\documentstyle[12pt]{article}
\begin{document}
\baselineskip 0.6cm

\def\simgt{\mathrel{\lower2.5pt\vbox{\lineskip=0pt\baselineskip=0pt
           \hbox{$>$}\hbox{$\sim$}}}}
\def\simlt{\mathrel{\lower2.5pt\vbox{\lineskip=0pt\baselineskip=0pt
           \hbox{$<$}\hbox{$\sim$}}}}
\def\lrD{\stackrel{\leftrightarrow}{\partial}}

\begin{titlepage}

\begin{flushright}
UCB-PTH-05/28 \\
LBNL-58858
\end{flushright}

\vskip 2.0cm

\begin{center}

{\Large \bf 
{\boldmath $\mu B$}-driven Electroweak Symmetry Breaking}

\vskip 1.0cm

{\large
Yasunori Nomura, David Poland, and Brock Tweedie
}

\vskip 0.4cm

{\it Department of Physics, University of California,
           Berkeley, CA 94720} \\
{\it Theoretical Physics Group, Lawrence Berkeley National Laboratory,
           Berkeley, CA 94720} \\

\vskip 1.2cm

\abstract{We consider a scenario in which the dominant quartic coupling 
for the Higgs doublets arises from the $F$-term potential, rather 
than the conventional $SU(2)_L \times U(1)_Y$ $D$-term potential, in 
supersymmetric theories.  The quartic coupling arises from a superpotential 
interaction between the two Higgs doublets and a singlet field, but 
unlike the case in the next-to-minimal supersymmetric standard model the 
singlet field is not responsible for the generation of the supersymmetric 
or holomorphic supersymmetry-breaking masses for the Higgs doublets.  
We find that this naturally leads to a deviation from the conventional 
picture of top-Yukawa driven electroweak symmetry breaking --- 
electroweak symmetry breaking is triggered by the holomorphic 
supersymmetry breaking mass for the Higgs doublets (the $\mu B$ term). 
This allows a significant improvement for fine-tuning in electroweak 
symmetry breaking, since the top squarks do not play a major role in 
raising the Higgs boson mass or in triggering electroweak symmetry 
breaking and thus can be light.  The amount of fine-tuning is given by 
the squared ratio of the lightest Higgs boson mass to the charged Higgs 
boson mass, which can be made better than $20\%$.  Solid implications 
of the scenario include a small value for $\tan\beta$, $\tan\beta 
\simlt 3$, and relatively light top squarks.}

\end{center}
\end{titlepage}

\section{Introduction}
\label{sec:intro}

One of the most pressing issues in particle theory is to understand the 
origin of electroweak symmetry breaking.  In the standard model, the 
effect of electroweak symmetry breaking is parameterized by the vacuum 
expectation value (VEV) of the Higgs field $h$, which is generated by 
the potential
\begin{equation}
  V = m_h^2\, |h|^2 + \frac{\lambda_h}{4} |h|^4,
\label{eq:higgs-pot}
\end{equation}
where $m_h^2$ is negative and $\lambda_h$ is positive.  Since $m_h^2$ 
is highly sensitive to the physics at the cutoff scale $\Lambda$, 
however, the standard model cannot adequately describe the origin of 
electroweak symmetry breaking --- the physics of electroweak symmetry 
breaking is at or beyond $\Lambda$. 

Weak scale supersymmetry provides an elegant framework to address this 
issue.  Since quadratic divergences to the Higgs boson mass-squared 
parameters are cut off by masses of superparticles, the physics of 
electroweak symmetry breaking can be described reliably within the 
framework of softly broken supersymmetric effective field theory. 
However, supersymmetry itself does not tell us the origin of the 
Higgs potential, Eq.~(\ref{eq:higgs-pot}).  Why is $m_h^2$ negative? 
What is the origin of $\lambda_h$?

The conventional answer to these questions, which has been dominating 
over 20 years, is that the restoring force in the potential, $\lambda_h$, 
arises from supersymmetric partners of weak interactions -- the $SU(2)_L 
\times U(1)_Y$ $D$-term potential -- while the mass-squared parameter, 
$m_h^2$, becomes negative because of the logarithmically divergent 
top-stop loop correction~\cite{Ibanez:1982fr}.  In fact, this provides 
an elegant answer to the most naive question in supersymmetric theories: 
why do the Higgs fields develop VEVs while squarks and sleptons do not? 
Since the origin of the instability is the top Yukawa coupling, the 
only fields that could get VEVs are the third generation squarks, 
$\tilde{q}_3$ and $\tilde{u}_3$, and the Higgs fields.  The famous 
color factor associated with the top-stop loop and/or a small hierarchy 
between colored and non-colored scalar masses can then explain why 
the Higgs fields are the only ones that obtain VEVs.

The framework just described, however, has been put in a subtle 
position after the negative results for the Higgs boson search at 
the LEP~II.  To evade the LEP~II bound on the physical Higgs boson 
mass, $M_{\rm Higgs} \simgt 114~{\rm GeV}$~\cite{Barate:2003sz}, the 
coupling $\lambda_h$ must satisfy $\lambda_h \simgt 0.43$ because 
the physical Higgs boson mass is given by
\begin{equation}
  M_{\rm Higgs}^2 = \lambda_h v^2,
\label{eq:higgs-mass}
\end{equation}
where $v \equiv \langle h \rangle \simeq 174~{\rm GeV}$.  In the 
minimal supersymmetric extension of the standard model (MSSM), the 
$SU(2)_L \times U(1)_Y$ $D$-terms give the quartic couplings for 
the Higgs fields $V = (g^2+g'^2)(|H_u^0|^2-|H_d^0|^2)^2/8$, where 
$g \simeq 0.65$ and $g' \simeq 0.36$ are the $SU(2)_L$ and $U(1)_Y$ 
gauge couplings, and $H_u$ and $H_d$ are the two Higgs doublets giving 
masses for the up-type and down-type quarks with the superscript 
$0$ denoting the neutral components.  Consider now that the field 
developing a VEV resides in $H_u$ and $H_d$ as $H_u^0 = h^\dagger 
\sin\beta$ and $H_d^0 = h \cos\beta$, where $\tan\beta \equiv \langle 
H_u^0 \rangle/\langle H_d^0 \rangle$.  This is a valid approximation 
if the Higgs bosons arising from the other linear combination are 
heavier than the excitation of $h$.  In this case we find
\begin{equation}
  \lambda_h = \frac{g^2+g'^2}{2} \cos^2\!2\beta.
\label{eq:mssm}
\end{equation}
This is too small to give $M_{\rm Higgs} \simgt 114~{\rm GeV}$. 
In the MSSM, we avoid this difficulty by invoking a radiative 
correction from top-stop loop, which gives an extra contribution to 
$\lambda_h$~\cite{Okada:1990vk}.  To evade the LEP~II bound, however, 
we need a rather large mass for the top squarks: $m_{\tilde{t}} \simgt 
(800\!\sim\!1200)~{\rm GeV}$ for a reasonably small top squark mixing 
parameter~\cite{Carena:1995wu}.  This in turn gives a large negative 
contribution to the $h$ mass-squared parameter
\begin{equation}
  \delta m_h^2 \simeq -\frac{3m_t^2}{4\pi^2 v^2}\, m_{\tilde{t}}^2\, 
    \ln\Biggl( \frac{M_{\rm mess}}{m_{\tilde{t}}} \Biggr),
\label{eq:corr-Higgs}
\end{equation}
where $M_{\rm mess}$ is the scale at which the squark and slepton 
masses are generated.  To reproduce the correct scale for electroweak 
symmetry breaking, we then need to cancel this with other positive 
contributions to $m_h^2$, for example the one from the supersymmetric 
mass for the Higgs fields.  In many supersymmetric models, this requires 
a cancellation of order a few percent or worse.  This problem is often 
called the supersymmetric fine-tuning problem.

The simplest way to avoid the supersymmetric fine-tuning problem is 
to make both $m_{\tilde{t}}^2$ and $\ln(M_{\rm mess}/m_{\tilde{t}})$ 
small.  Suppose that $\ln(M_{\rm mess}/m_{\tilde{t}})$ is as 
small as a factor of a few, which is reasonable if the mediation 
scale of supersymmetry breaking is low.  In this case the 
fine-tuning can be made better than $(20\!\sim\!30)\%$ if 
$m_{\tilde{t}} \simlt (400\!\sim\!510)(M_{\rm Higgs}/150~{\rm GeV})
(3.5/\ln(M_{\rm mess}/m_{\tilde{t}}))^{1/2}$~\cite{Chacko:2005ra,%
Nomura:2005qg}.  However, such light top squark masses require some 
ingredient to be reconciled with the LEP~II bound of $M_{\rm Higgs} 
\simgt 114~{\rm GeV}$.  One of the simplest possibilities then is 
to introduce an additional contribution to the Higgs quartic coupling 
other than the $SU(2)_L \times U(1)_Y$ $D$-terms and the top-stop loop.%
\footnote{An alternative possibility is to have a large trilinear 
scalar coupling for the top squarks.  A solution to the supersymmetric 
fine-tuning problem using this has been presented in~\cite{Kitano:2005wc}. 
The effect of the trilinear coupling on fine-tuning was also analyzed, 
e.g., in~\cite{Kane:2003iq}.}
Suppose now that we actually have such a contribution.  We then find 
that, unless it comes from the coupling of the form $|H_u|^4$, 
the contribution from the $D$-terms, Eq.~(\ref{eq:corr-Higgs}), 
is suppressed, because any other forms of couplings, such as 
$|H_u H_d|^2$, give powers of $\cos\beta$ when expressed in terms 
of $h$, so that $\tan\beta$ is required not to be very large (i.e. 
$\cos^2\!2\beta$ is appreciably smaller than $1$) for the new 
contribution to be effective.  This suggests an alternative picture 
for the origin of the Higgs potential in supersymmetric theories: 
the dominant source of the Higgs quartic coupling may not be the 
$SU(2)_L \times U(1)_Y$ $D$-terms. 

In this paper we study a framework in which the {\it dominant} 
contribution to $\lambda_h$ comes from the $F$-term potential for 
the Higgs fields, rather than the $D$-terms of $SU(2)_L \times U(1)_Y$. 
While such a contribution arises from coupling the Higgs fields with 
some other field (a gauge singlet field) in the superpotential, we are 
not particularly interested in the dynamics of the additional field. 
In fact, we mainly consider the cases in which the dynamics of this 
field essentially decouples.  We find that this naturally leads us 
to consider a scenario in which not only the origin of $\lambda_h$ 
but also the reason for the instability, $m_h^2 < 0$, deviate from 
the conventional picture.  In this scenario, it is the presence of 
the holomorphic supersymmetry breaking mass term ($\mu B$ term) for 
the Higgs fields that is crucial for electroweak symmetry breaking. 
The top-stop loop contribution of Eq.~(\ref{eq:corr-Higgs}) also 
plays a role, but in the absence of the $\mu B$ term the electroweak 
symmetry is not broken.  In the conventional picture, the top-stop 
loop itself drives the squared mass of the (up-type) Higgs field 
negative, triggering electroweak symmetry breaking.  The role of 
the $\mu B$ term is then to make $\tan\beta$ differ from infinity. 
In our picture the electroweak symmetry is broken by the presence 
of the $\mu B$ term.  The role of the top-stop loop contribution 
then is, besides contributing to electroweak symmetry breaking, to 
make $\tan\beta$ deviate from $1$, which would be the case if the 
squared masses for the up-type and down-type Higgs fields were 
the same.

One of the major advantages of the present scenario over the 
conventional one is that the masses of the top squarks can be small 
as they do not play a major role in raising the Higgs boson mass nor 
in triggering electroweak symmetry breaking.  This makes it possible 
to evade severe fine-tuning in electroweak symmetry breaking.  We find 
that the fine-tuning in our scenario is determined essentially by the 
size of the $\mu B$ parameter, which in turn is related to the mass 
of the charged Higgs boson.  We find that fine-tuning can be made 
better than $20\%$ for reasonable values of the lightest Higgs boson 
mass and the charged Higgs boson mass.

We note here that similar dynamics for the Higgs sector were considered 
in Ref.~\cite{Brignole:2003cm}.  There the possibility of electroweak 
symmetry breaking by the $\mu B$ term and the reduction of fine-tuning 
by large quartic couplings were discussed for generic Higgs potentials 
in the context of very low ($O({\rm TeV})$) scale fundamental 
supersymmetry breaking.

In the next section we discuss our scenario.  The basic framework 
is described in section~\ref{subsec:framework}, and the two 
particularly interesting cases are studied in more detail in 
sections~\ref{subsec:large-mS2} and \ref{subsec:large-MS}.  The 
simple formulae for the fine-tuning are derived.  Some implications 
of the scenario are discussed in~\ref{subsec:impli}.  Conclusions 
are given in section~\ref{sec:concl}.

\section{{\boldmath $\mu B$}-driven Electroweak Symmetry Breaking}
\label{sec:ewsb}

\subsection{The basic framework}
\label{subsec:framework}

Let us consider the superpotential term
\begin{equation}
  W = \lambda S H_u H_d,
\label{eq:SHH}
\end{equation}
where $S$ is a gauge singlet field and $\lambda$ is a dimensionless 
coupling.  The $SU(2)_L$ contraction for the Higgs fields is defined 
by $H_u H_d \equiv H_u^+ H_d^- - H_u^0 H_d^0$.  This is the simplest 
term that can give an $F$-term contribution to the Higgs quartic 
coupling, without destroying the successful supersymmetric prediction 
for gauge coupling unification~\cite{Dimopoulos:1981zb}.  The value 
of the coupling $\lambda$ at the weak scale is constrained by a 
Landau pole consideration.  Requiring that $\lambda$, as well as 
the gauge and Yukawa couplings, are perturbative up to the unification 
scale of $10^{16}~{\rm GeV}$, the weak scale value of $\lambda$ is 
constrained to be $\lambda \simlt (0.6\!\sim\!0.8)$, depending on 
the value of $\tan\beta$ and the assumption on the field content 
between the weak and the unification scales~\cite{Masip:1998jc}. 
A larger value of $\lambda$, however, may be obtained if some of 
the $S$, $H_u$ and $H_d$ fields are composite states arising at low 
scales~\cite{Harnik:2003rs} or if the gauge group is extended at 
the weak scale~\cite{Batra:2004vc}.  As we will see explicitly later, 
these values of $\lambda$ can produce a stabilizing force in the Higgs 
potential (the quartic coupling) stronger than that arising from the 
$SU(2)_L \times U(1)_Y$ $D$-terms.

Unlike the case for the next-to-minimal supersymmetric standard 
model (NMSSM)~\cite{Nilles:1982dy}, we are not interested in using 
the interaction of Eq.~(\ref{eq:SHH}) to explain the origin of the 
supersymmetric mass term ($\mu$ term) for the Higgs doublets.%
\footnote{For discussions of fine-tuning in the NMSSM in various 
contexts, see e.g.~[\ref{Chacko:2005ra:X},~\ref{Agashe:1997kn:X}%
~--~\ref{Dermisek:2005ar:X}].}
Rather, we consider that the $\mu$ term, as well as the $\mu B$ 
term, arise from some other source.  Such a setup can naturally 
arise, for example, if some of the $S$, $H_u$ and $H_d$ fields are 
coupled to the sector that breaks supersymmetry dynamically at a 
scale of order $(10\!\sim\!100)~{\rm TeV}$ through nearly marginal 
operators~\cite{Chacko:2005ra,Goldberger:2002pc}.  Alternatively, 
the $\mu$ and $\mu B$ terms may arise from supergravity 
effects~\cite{Giudice:1988yz} if the fundamental scale of 
supersymmetry breaking is at an intermediate scale~\cite{NPT1}. 

The Higgs sector of our framework is then specified by the following 
superpotential and soft supersymmetry breaking Lagrangian:
\begin{equation}
  W = \lambda S H_u H_d + \mu H_u H_d + \hat{W}(S),
\label{eq:superpot}
\end{equation}
and
\begin{equation}
  {\cal L}_{\rm soft} = - m_{H_u}^2 |H_u|^2 
    - m_{H_d}^2 |H_d|^2 - (\mu B\, H_u H_d + {\rm h.c.}) 
    + \hat{\cal L}_{\rm soft}(S,H_u,H_d),
\label{eq:soft-Lag}
\end{equation}
where $\hat{W}(S)$ is a holomorphic function of $S$, and 
$\hat{\cal L}_{\rm soft}(S,H_u,H_d)$ represents soft supersymmetry 
breaking terms that contain the $S$ field.  In addition, the Higgs 
potential contains the terms arising from the $SU(2)_L \times U(1)_Y$ 
$D$-terms, which we write as 
\begin{equation}
  V_D = \epsilon \frac{g^2}{2} \sum_{a=1}^{3} 
    \biggl( H_u^\dagger \frac{\sigma^a}{2} H_u 
      + H_d^\dagger \frac{\sigma^a}{2} H_d \biggr)^2
    + \epsilon' \frac{g'^2}{2} \biggl( \frac{1}{2} H_u^\dagger H_u 
      - \frac{1}{2} H_d^\dagger H_d \biggr)^2,
\label{eq:D-pot}
\end{equation}
where $\sigma^a$ are the Pauli matrices.  Here, we have introduced 
the factors $\epsilon$ and $\epsilon'$, which take values $\epsilon, 
\, \epsilon' \leq 1$, to take into account potential suppressions 
of these $D$-terms.  Such suppressions arise, for example, 
in theories in which the gauginos obtain Dirac masses though 
supersymmetry breaking operators coupling them with other adjoint 
fermions~\cite{Fox:2002bu}.  Our electroweak symmetry breaking 
scenario works even in the case with $\epsilon, \, \epsilon' \ll 1$, 
since the existence of the $D$-term potential is not important. 
In the absence of the Dirac gaugino masses, as in the usual case, 
$\epsilon = \epsilon' = 1$.  We assume that the parameters appearing 
in Eqs.~(\ref{eq:superpot},~\ref{eq:soft-Lag}) are real throughout 
the paper. 

The fact that we do not need to generate the $\mu$ or $\mu B$ term 
from the interaction of Eq.~(\ref{eq:SHH}) allows considerable 
freedoms for $\hat{W}(S)$ and $\hat{\cal L}_{\rm soft}(S,H_u,H_d)$. 
Any constraints that would arise from requiring the appreciable 
size of the $S$ VEV to be generated are absent here.  In fact, this 
allows very simple forms for $\hat{W}(S)$ and $\hat{\cal L}_{\rm 
soft}(S,H_u,H_d)$ as realistic possibilities.  In this paper we 
particularly focus on the cases in which the dynamics of the $S$ 
field almost decouples from the physics of electroweak symmetry 
breaking and the VEV of $S$ is much smaller than the weak scale. 
The purpose of the interaction of Eq.~(\ref{eq:SHH}), then, is to 
provide the source of the Higgs quartic coupling, $\lambda_h$, needed 
to push up the Higgs boson mass.  (This interaction also modifies 
the spectrum in the Higgs sector by mixing the $H_u$ and $H_d$ 
states with the $S$ states.)

As we will see later, the quartic coupling arising from the 
interaction of Eq.~(\ref{eq:SHH}) takes the form of either 
$|H_u H_d|^2$ or $(|H_u|^2+|H_d|^2)(H_u H_d + {\rm h.c.})$, giving 
a contribution to $\lambda_h$ proportional to $\sin^2\!2\beta$ or 
$\sin2\beta$.  This implies that these $F$-term quartic couplings 
do not stabilize the direction $|H_u| \rightarrow \infty$ with 
$H_d = 0$ or $|H_d| \rightarrow \infty$ with $H_u = 0$ in field 
space.  Of course, these directions are eventually stabilized by 
the quartic couplings arising from the $SU(2)_L \times U(1)_Y$ 
$D$-term potential, Eq.~(\ref{eq:D-pot}), but then we are relying on 
the weak stabilization force from the gauge $D$-terms and the situation 
is not very much different from the conventional scenario.  Therefore, 
to use a stronger $F$-term quartic coupling as the dominant restoring 
force, we consider the situation in which the directions $H_u H_d = 0$ 
are stabilized by the {\it quadratic} terms in the potential.  This 
requires that the squared masses for {\it both} the up-type and 
down-type Higgs fields are positive:
\begin{equation}
  |\mu|^2 + m_{H_u}^2 > 0 \qquad {\rm and} \qquad 
  |\mu|^2 + m_{H_d}^2 > 0.
\label{eq:Higgs-diag}
\end{equation}
Electroweak symmetry breaking is then necessarily caused by the 
$\mu B$ term which satisfies
\begin{equation}
  |\mu B|^2 > (|\mu|^2 + m_{H_u}^2)(|\mu|^2 + m_{H_d}^2),
\label{eq:Higgs-offdiag}
\end{equation}
by making one of the eigenvalues in the Higgs mass-squared matrix 
negative.  For this reason, we call the scenario presented here 
{\it $\mu B$-driven electroweak symmetry breaking}.

In the present scenario, the role of the negative top-stop loop 
contribution to $m_{H_u}^2$ is, besides facilitating to satisfy 
Eq.~(\ref{eq:Higgs-offdiag}), to deviate $\tan\beta$ from $1$. 
If this contribution were absent, the squared masses for the up-type 
and down-type Higgs fields would be the same so that $\tan\beta$ 
would exactly be $1$, because the theory then possesses a symmetry 
$H_u \leftrightarrow H_d^\dagger$ (in the limit of neglecting all 
the other Yukawa couplings).%
\footnote{In general two Higgs doublet potentials, a symmetry 
$H_u \leftrightarrow H_d^\dagger$ can be spontaneously broken, 
giving $\tan\beta \neq 1$~\cite{Brignole:2003cm}.  This does 
not happen in our case because of the supersymmetric relations 
between the quartic couplings.}
In fact, $\tan\beta$ in this scenario is given by $\tan^2\!\beta 
\approx (|\mu|^2 + m_{H_d}^2)/(|\mu|^2 + m_{H_u}^2)$, explicitly 
showing that $m_{H_u}^2 \neq m_{H_d}^2$ is the source of $\tan\beta 
\neq 1$.  This equation also tells us that $\tan\beta$ is expected to 
be not very large in the present scenario, e.g. $\tan\beta \simlt 3$, 
because a very large $\tan\beta$ requires a fine cancellation between 
the two terms in the denominator of the right-hand-side.  This is 
consistent with the fact that we use the quartic coupling proportional 
to $\sin^2\!2\beta$ or $\sin2\beta$, which would vanish in the limit 
of large $\tan\beta$.

We note that the scenario described here provides an alternative 
answer to the question: why do only the Higgs fields develop VEVs? 
The answer is: because they are the only fields that are vector-like 
under the standard model gauge group.  We have seen that electroweak 
symmetry breaking is triggered by the holomorphic supersymmetry 
breaking mass term, the $\mu B$ term, in the Higgs sector.  For the 
squarks and sleptons we cannot write down such terms because they 
belong to supermultiplets that are chiral under the standard model 
gauge group.  Analogous terms can arise after electroweak symmetry 
breaking (the left-right mixing term), but they are generically 
smaller.  For the top squarks, however, we may need an additional 
argument, such as the one that colored particles are somewhat heavier 
than the others, depending on the model of supersymmetry breaking.

\subsection{Scenario~I --- Supersymmetric Higgs with a 
``non-supersymmetric'' quartic coupling}
\label{subsec:large-mS2}

We now want to discuss the dynamics of electroweak symmetry breaking 
in more detail.  For this purpose we can integrate out the $S$ field 
at tree level and obtain the effective theory containing only the 
Higgs doublet fields.  In general, the functions $\hat{W}(S)$ and 
$\hat{\cal L}_{\rm soft}(S,H_u,H_d)$ contain arbitrary mass parameters 
of order the weak scale.  The integration of $S$ in the general setup 
is performed in the Appendix. 

We are particularly interested in the situation in which the dynamics 
of $S$ essentially decouples.  We here consider the case in which the 
decoupling occurs because of a large soft supersymmetry breaking mass 
of $S$: $m_S^2 \gg v^2, |M_S|^2, |b_S|$, where $M_S$ and $b_S$ are 
the supersymmetric mass and the holomorphic supersymmetry breaking mass 
squared for $S$, respectively.  By large, we mean $(m_S^2)^{1/2}$ of the 
order of several hundreds of GeV.  Since $m_S^2$ gives a contribution 
to $m_{H_u}^2$ at the one-loop level, it cannot be larger than about 
$(8\pi^2/\lambda^2)v^2$.

The effective Higgs potential for the case with $m_S^2 \gg v^2$ is 
obtained simply by setting $S = 0$ in the potential derived from 
Eqs.~(\ref{eq:superpot},~\ref{eq:soft-Lag},~\ref{eq:D-pot}), assuming 
that $\hat{\cal L}_{\rm soft}(S,H_u,H_d)$ does not contain a large 
linear $S$ term.  In the absence of the linear term in $\hat{W}(S)$, 
the potential for the two Higgs doublets is given by
\begin{equation}
  V = |\lambda H_u H_d|^2 + (|\mu|^2 + m_{H_u}^2) |H_u|^2 
      + (|\mu|^2 + m_{H_d}^2) |H_d|^2 + (\mu B\, H_u H_d + {\rm h.c.}) 
      + V_D.
\label{eq:pot-model-1}
\end{equation}
(If $\hat{W}(S)$ contains the linear term $L_S^2 S$, $\mu B$ should 
be redefined as $\mu B \rightarrow \mu B - \lambda L_S^2$ to absorb 
the effect of that term.%
\footnote{In addition, if $\hat{\cal L}_{\rm soft}(S,H_u,H_d)$ contains 
a large linear $S$ term, $\hat{\cal L}_{\rm soft} = -c_S S + {\rm h.c.}$ 
with $c_S \approx m_S^2 v$, both $\mu$ and $\mu B$ should be redefined 
to absorb the linear terms: $\mu \rightarrow \mu + \lambda c_S/m_S^2$ 
and $\mu B \rightarrow \mu B - \lambda L_S^2 + \lambda c_S M_S/m_S^2 
- \lambda \kappa c_S^2/m_S^4 + a_\lambda c_S/m_S^2$, where $\kappa$ 
and $a_\lambda$ are the coefficients of the $S^3/3$ and $-S H_u 
H_d$ terms in $\hat{W}$ and $\hat{\cal L}_{\rm soft}$, respectively. 
The appropriate redefinitions for the origin of $S$ and parameters 
in $\hat{W}$ and $\hat{\cal L}_{\rm soft}$ are also needed.  These 
redefinitions can generate $\mu$ and $\mu B$ terms even if they are 
absent before the redefinitions.}%
)  This potential can, in fact, be obtained by taking the limit 
$m_S^2 \gg |M_S|^2, |b_S|, |\mu|^2, |a_\lambda|^2$ in the expression 
in the Appendix.  We find that the potential takes the same form 
as the MSSM potential except that there is a new quartic term, 
$|\lambda H_u H_d|^2$.  Since the $S$ scalar is relatively heavy, it 
effectively appears at the scale $v$ that a ``non-supersymmetric'' 
quartic coupling has been added to the supersymmetric Higgs 
potential~\cite{Espinosa:1998re}.

Approximating that $H_u^0 = h^\dagger \sin\beta$ and $H_d^0 = h \cos\beta$, 
the Higgs quartic coupling of Eq.~(\ref{eq:higgs-pot}) is given by
\begin{equation}
  \lambda_h = \lambda^2 \sin^2\!2\beta 
    + \frac{\epsilon g^2 + \epsilon' g'^2}{2} \cos^2\!2\beta.
\label{eq:quartic-model-1}
\end{equation}
The lightest Higgs boson mass is then given by Eq.~(\ref{eq:higgs-mass}). 
This is a good approximation as long as the other Higgs bosons are 
somewhat heavier, i.e. $2|\mu|^2 + m_{H_u}^2 + m_{H_d}^2 \simgt M_{\rm 
Higgs}^2$, which we expect to be the case.  Our scenario assumes 
that the dominant contribution to $\lambda_h$ is the first term in 
Eq.~(\ref{eq:quartic-model-1}).  This is most effective if $\lambda^2 
\sin^2\!2\beta$ is larger than the largest tree-level MSSM value, 
$(g^2 + g'^2)/2$, which occurs if $\lambda \simgt 0.6$ ($0.7$) for 
$\tan\beta \simeq 1.7$ ($2.2$).  (In the case of $\epsilon, \, \epsilon' 
\ll 1$, this becomes a necessary condition to reduce fine-tuning.) 
Such values of $\lambda$ require deviations from the simplest MSSM 
at energies above the weak scale, for example, stronger gauge couplings 
at higher energies~\cite{Chacko:2005ra,Masip:1998jc}.

The minimization of the potential, Eq.~(\ref{eq:pot-model-1}), gives
\begin{eqnarray}
  && |\mu|^2 + m_{H_u}^2 = \mu B \frac{1}{\tan\beta} - v^2 
      \left\{ \lambda^2 \cos^2\!\beta + \frac{\epsilon g^2 
      + \epsilon' g'^2}{4} (\sin^2\!\beta - \cos^2\!\beta) \right\},
\label{eq:minimize-1}
\\
  && |\mu|^2 + m_{H_d}^2 = \mu B \tan\beta - v^2 
      \left\{ \lambda^2 \sin^2\!\beta + \frac{\epsilon g^2 
      + \epsilon' g'^2}{4} (\cos^2\!\beta - \sin^2\!\beta) \right\}.
\label{eq:minimize-2}
\end{eqnarray}
It is customary to rewrite these equations in the form
\begin{eqnarray}
  && \frac{\epsilon g^2 + \epsilon' g'^2}{4} v^2 
    = \frac{m_{H_d}^2 - m_{H_u}^2 \tan^2\!\beta}{\tan^2\!\beta-1} 
      - |\mu|^2,
\label{eq:min-wrong-1}
\\
  && 2 |\mu|^2 + m_{H_u}^2 + m_{H_d}^2 = \frac{2\mu B}{\sin 2 \beta} 
      - \lambda^2 v^2.
\label{eq:min-wrong-2}
\end{eqnarray}
In the corresponding equations in the MSSM, the $\lambda^2 v^2$ term 
in the second equation is absent (and $\epsilon = \epsilon' = 1$ in the 
first equation).  In this case the second equation determines $\tan\beta$ 
and then the electroweak scale $v$ is determined by the first equation. 
In our case, however, both $v$ and $\beta$ appear in the two equations. 
It is therefore not obvious that the ``roles'' of these two equations 
are the same as those in the MSSM.  In fact, in our scenario where 
the dominant restoring force arises from the $F$-term potential, it 
is the second equation that determines $v$.  This is easily understood 
if we take the limit $g,\, g' \rightarrow 0$ in Eq.~(\ref{eq:min-wrong-1}) 
(since they are supposed to play negligible roles) --- the first equation 
determines $\tan\beta$ and then $v$ is determined by the second equation.

We find it useful to rewrite Eqs.~(\ref{eq:min-wrong-1},~%
\ref{eq:min-wrong-2}) in the form
\begin{eqnarray}
  && \tan^2\!\beta = \frac{|\mu|^2 + m_{H_d}^2 + \hat{m}_Z^2/2}
      {|\mu|^2 + m_{H_u}^2 + \hat{m}_Z^2/2},
\label{eq:min-correct-1}
\\
  && \lambda^2 v^2 = \frac{2\mu B}{\sin 2 \beta} 
    - (2 |\mu|^2 + m_{H_u}^2 + m_{H_d}^2),
\label{eq:min-correct-2}
\end{eqnarray}
where $\hat{m}_Z^2 \equiv (\epsilon g^2 + \epsilon' g'^2) v^2/2$, 
which becomes the $Z$-boson mass squared for $\epsilon=\epsilon'= 1$. 
This makes it manifest that, in the parameter region of interest 
$|\mu|^2, |m_{H_u}^2|, |m_{H_d}^2| \simgt \hat{m}_Z^2/2$, $\tan\beta$ 
is mostly determined by the first equation while $v$ by the second 
equation.  Equation~(\ref{eq:min-correct-1}) also makes it obvious 
that, as was observed in the previous general discussion, the 
significant fine-tuning between $|\mu|^2$ and $m_{H_u}^2$ can be 
avoided for $|\mu|^2 + m_{H_u}^2 > 0$, and in that case $\tan\beta 
\approx ((|\mu|^2 + m_{H_d}^2)/(|\mu|^2 + m_{H_u}^2))^{1/2}$ is 
expected to be not large.  

The electroweak scale, $v$, is determined by the second equation, 
Eq.~(\ref{eq:min-correct-2}).  In the absence of a tuning between 
$|\mu|^2$ and $m_{H_u}^2$, the only source of a potential tuning 
is in the right-hand-side of this equation.  Now, the combination 
$2|\mu|^2 + m_{H_u}^2 + m_{H_d}^2$ appearing in the right-hand-side 
is related to the charged Higgs boson mass, $M_{H^{\pm}}$, as 
\begin{equation}
  M_{H^{\pm}}^2 = 2|\mu|^2 + m_{H_u}^2 + m_{H_d}^2 + \hat{m}_W^2,
\label{eq:Hpm}
\end{equation}
where $\hat{m}_W^2 \equiv \epsilon g^2 v^2/2$, which becomes the 
$W$-boson mass squared for $\epsilon= 1$.  To evade the constraint 
from the $b \rightarrow s \gamma$ process without relying on 
excessive cancellations between different contributions, $M_{H^{\pm}}$ 
is expected to be somewhat larger than $v$, e.g. $M_{H^{\pm}} \simgt 
250~{\rm GeV}$~\cite{Borzumati:1998tg}.  This, therefore, requires 
some cancellation between the first term and the rest in the 
right-hand-side of Eq.~(\ref{eq:min-correct-2}).  The amount of the 
required cancellation is given by the following fine-tuning parameter:
\begin{equation}
  \Delta^{-1} \simeq \frac{\lambda^2 v^2}{M_{H^{\pm}}^2-\hat{m}_W^2}
    \approx \frac{M_{\rm Higgs}^2}{M_{H^{\pm}}^2},
\label{eq:fine-tuning-1}
\end{equation}
where the last relation holds approximately for $\tan\beta \simlt 2$. 
Consider, for example, the case with $\lambda \simeq 0.8$, which can 
be obtained consistently with the Landau pole bound without relying 
on the compositeness of $S$, $H_u$ or $H_d$.  In this case, the 
requirement $\Delta^{-1} > 20\%$ ($10\%$) gives $M_{H^{\pm}} \simlt 
320~{\rm GeV}$ ($450~{\rm GeV}$) for $\epsilon = \epsilon' = 1$ and 
$M_{H^{\pm}} \simlt 310~{\rm GeV}$ ($440~{\rm GeV}$) for $\epsilon = 
\epsilon' = 0$.  For $\lambda \simeq 0.7$, the corresponding bound 
becomes $M_{H^{\pm}} \simlt 280~{\rm GeV}$ ($390~{\rm GeV}$). 
For these values of $M_{H^{\pm}}$, some amount of a destructive 
interference between the chargino and charged Higgs boson 
contributions in the $b \rightarrow s \gamma$ amplitude may be 
needed, and this prefers the positive sign for the $\mu$ parameter. 

In the above analysis, the terms $\hat{m}_Z^2/2$ in 
Eq.~(\ref{eq:min-correct-1}) were not important.  This clearly 
shows that the $SU(2)_L \times U(1)_Y$ $D$-term potential is not 
playing a major role in electroweak symmetry breaking, although it gives 
a small correction to the Higgs boson mass through the second term of 
Eq.~(\ref{eq:quartic-model-1}).  In fact, this was the dynamics behind 
the reduction of fine-tuning in the minimally fine-tuned model of 
Ref.~\cite{Chacko:2005ra}, in which the necessary $\mu$ and $\mu B$ 
terms arise effectively from the linear $S$ terms in $\hat{W}(S)$ 
and $\hat{\cal L}_{\rm soft}(S,H_u,H_d)$.  One can check that our 
$\Delta^{-1}$ in Eq.~(\ref{eq:fine-tuning-1}) well reproduces the 
values of the fine-tuning parameter $\tilde{\Delta}^{-1}$, defined 
in Ref.~\cite{Chacko:2005ra} following~\cite{Anderson:1994dz}. 

At the leading order in the $1/m_S^2$ expansion, the lightest 
Higgs boson mass is approximately given by Eq.~(\ref{eq:higgs-mass}) 
with Eq.~(\ref{eq:quartic-model-1}), and the charged Higgs boson mass 
is given by Eq.~(\ref{eq:Hpm}).  The masses for the heavier neutral 
Higgs boson and the pseudo-scalar Higgs boson are given by $M_{H^0}^2 
\simeq 2|\mu|^2 + m_{H_u}^2 + m_{H_d}^2 + \lambda^2 v^2 \cos^2\!2\beta 
+ \hat{m}_Z^2 \sin^2\!2\beta$ and $M_{A^0}^2 = 2|\mu|^2 + m_{H_u}^2 
+ m_{H_d}^2 + \lambda^2 v^2$, respectively.%
\footnote{The precise formulae for the masses of the lightest and 
heavier Higgs bosons in the $S$ decoupling limit are $M_{h,H^0}^2 
= [M_{A^0}^2+\hat{m}_Z^2 \pm \{ (M_{A^0}^2-\hat{m}_Z^2)^2 \cos^2\!2\beta 
+ (M_{A^0}^2+\hat{m}_Z^2-2 \lambda^2 v^2)^2 \sin^2\!2\beta \}^{1/2}]/2$.}
These masses, however, are modified by the mixings with the $S$ states 
except for the mass of the charged Higgs boson.  The mixing effects 
can be taken into account if we use the effective Lagrangian derived 
using the complete $S$ propagators given in the Appendix.  Many of the 
effects, in fact, can be captured even in the effective potential of 
Eq.~(\ref{eq:app-pot}) derived using the zero-momentum $S$ propagators. 
These effects typically cause shifts of the masses at the level of $10\%$.

\subsection{Scenario~II --- Supersymmetric Higgs with a little 
seesaw at sub-TeV}
\label{subsec:large-MS}

We now consider an alternative scenario in which the dynamics of 
$S$ decouples because of a large supersymmetric mass, $M_S \gg v, 
(m_S^2)^{1/2}, |b_S|^{1/2}$.  Here, by large we mean that $M_S$ 
is only somewhat larger than $v$, of the order of several hundreds 
of GeV.  We assume that the linear $S$ terms in $\hat{W}(S)$ and 
$\hat{\cal L}_{\rm soft}(S,H_u,H_d)$ are absent, or their effects 
are already absorbed in the appropriate redefinitions of parameters.

For $M_S \gg (m_S^2)^{1/2}, |b_S|^{1/2}$, we can supersymmetrically 
integrate out the $S$ multiplet and obtain the effective superpotential 
and the K\"ahler potential for the two Higgs doublets, describing the 
physics at the scale $v$.  For $\hat{W}(S) = (M_S/2) S^2$, the effective 
superpotential is given by
\begin{equation}
  W = \mu H_u H_d - \frac{\lambda^2}{2 M_S} (H_u H_d)^2,
\label{eq:eff-suppot}
\end{equation}
while the effective K\"ahler potential is given by
\begin{equation}
  K = |H_u|^2 + |H_d|^2 + \frac{\lambda^2}{M_S^2} |H_u H_d|^2,
\label{eq:eff-Kahler}
\end{equation}
where the last terms in these equations are obtained using the two 
supersymmetric terms $(M_S/2) S^2$ and $\lambda S H_u H_d$ through a 
``little seesaw mechanism.''  We drop possible absolute value symbols 
for the parameters here and below, as they appear only for the 
squared quantities and all the parameters are assumed to be real. 

The dynamics of electroweak symmetry breaking is described by this 
superpotential and K\"ahler potential and the soft supersymmetry 
breaking Lagrangian of Eq.~(\ref{eq:soft-Lag}) with $\hat{\cal L}_{\rm 
soft}(S,H_u,H_d)$ set to zero.  The Higgs potential is then given by
\begin{eqnarray}
  V &=& \frac{\{ \mu^2 M_S^2 -\lambda^2 \mu M_S (H_u H_d+{\rm h.c.}) 
    + \lambda^4 |H_u H_d|^2 \} (|H_u|^2+|H_d|^2)}
    {M_S^2 + \lambda^2(|H_u|^2+|H_d|^2)}
\nonumber\\
  && {} + m_{H_u}^2 |H_u|^2 + m_{H_d}^2 |H_d|^2 
      + (\mu B\, H_u H_d + {\rm h.c.}) + V_D.
\label{eq:pot-model-2}
\end{eqnarray}
In the true $M_S \rightarrow \infty$ limit, the $F$-term potential 
in the first line simply gives supersymmetric masses, $\mu$, to $H_u$ 
and $H_d$.  The appreciable quartic couplings, however, arise if 
$\mu/M_S$ is not very small.  The potential of Eq.~(\ref{eq:pot-model-2}) 
can also be obtained from Eq.~(\ref{eq:app-pot}) in the Appendix 
by taking the limit $|M_S|^2 \gg m_S^2, |b_S|, |a_\lambda|^2$ with 
$|\mu| \simgt |a_\lambda/\lambda|$.

Approximating that $H_u^0 = h^\dagger \sin\beta$ and $H_d^0 = h \cos\beta$, 
the Higgs quartic coupling, $\lambda_h$, is given by
\begin{equation}
  \lambda_h = \frac{4 \lambda^2 \mu}{M_S} \sin 2\beta
    - \frac{4 \lambda^2 \mu^2}{M_S^2}
    + \frac{\epsilon g^2 + \epsilon' g'^2}{2} \cos^2\!2\beta.
\label{eq:quartic-model-2}
\end{equation}
We then find that the physics associated with the potential 
Eq.~(\ref{eq:pot-model-2}) depends strongly on the values of parameters. 
Let us take the convention that the sign of $\mu B$ is positive.  (This 
choice can always be made by rotating the phase of $H_u H_d$.)  The 
minimization conditions are given by
\begin{eqnarray}
  && \mu^2 + m_{H_u}^2 = \mu B \frac{1}{\tan\beta} - v^2 
      \left\{ \frac{\lambda^2 \mu}{M_S} \Bigl( \sin 2\beta 
      + \frac{1}{\tan\beta} \Bigr) - \frac{2 \lambda^2 \mu^2}{M_S^2} 
      - \frac{\epsilon g^2 + \epsilon' g'^2}{4} \cos 2\beta \right\},
\label{eq:minimize-1-Ms}
\\
  && \mu^2 + m_{H_d}^2 = \mu B \tan\beta - v^2 
      \left\{ \frac{\lambda^2 \mu}{M_S} \Bigl( \sin 2\beta 
      + \tan\beta \Bigr) - \frac{2 \lambda^2 \mu^2}{M_S^2} 
      + \frac{\epsilon g^2 + \epsilon' g'^2}{4} \cos 2\beta \right\}.
\label{eq:minimize-2-Ms}
\end{eqnarray}
Subtracting these two equations, we obtain
\begin{equation}
  \left( \mu B - \frac{\lambda^2 v^2 \mu}{M_S} \right) 
      \left( \tan\beta - \frac{1}{\tan\beta} \right)
    = m_{H_d}^2 - m_{H_u}^2 + \frac{\epsilon g^2 + \epsilon' g'^2}{2} 
      v^2 \cos 2\beta.
\label{eq:subtr}
\end{equation}
We assume that the right-hand-side of this equation is positive, which 
is always the case if the difference between $m_{H_u}^2$ and $m_{H_d}^2$ 
arises from the top-stop loop contribution of Eq.~(\ref{eq:corr-Higgs}) 
with $\ln(M_{\rm mess}/m_{\tilde{t}}) \simgt 1$ (and $m_{\tilde{t}} 
\simgt 300~{\rm GeV}$).  We then find that for $\mu B - \lambda^2 v^2 
\mu/M_S > 0$ ($< 0$), $\tan 2\beta$ is negative (positive).  Let us first 
consider the case with $\mu B - \lambda^2 v^2 \mu/M_S < 0$.  In this case, 
$\lambda^2 \mu/M_S$ is positive while $\sin 2\beta$ is negative for 
$|H_u| > |H_d|$, so from Eq.~(\ref{eq:quartic-model-2}) we find that the 
Higgs potential is not stabilized by the quartic coupling arising from 
the $F$-term.  This is not good, since then we have to rely on 6-point 
interactions to stabilize the VEVs, which are too weak to give a 
phenomenologically acceptable value for the Higgs boson mass.  We are 
thus led to consider the case with $\mu B - \lambda^2 v^2 \mu/M_S > 0$. 
In this case, the $F$-term contributions to the quartic coupling can 
provide a sufficiently strong stabilizing force, as long as the first 
term in Eq.~(\ref{eq:quartic-model-2}) is positive, i.e. $\lambda^2 
\mu/M_S$ is positive.  We thus finally obtain the condition for the 
present scenario to work:
\begin{equation}
  \mu B > \frac{\lambda^2 v^2 \mu}{M_S} > 0.
\label{eq:cond-model-2}
\end{equation}
With this condition satisfied, the effects from 6-point interactions 
are small in generic parameter regions, and the lightest Higgs 
boson mass is given approximately by Eq.~(\ref{eq:higgs-mass}) 
with Eq.~(\ref{eq:quartic-model-2}). 

Neglecting the contribution from the $D$-term in the right-hand-side, 
which is a reasonable approximation in most of the parameter space, 
Eq.~(\ref{eq:subtr}) can be used to determine the value of $\tan\beta$. 
For $\mu B$ sufficiently larger than $\lambda^2 v^2 \mu/M_S$, we find
\begin{equation}
  \tan 2\beta \approx -\frac{2 \mu B}{m_{H_d}^2 - m_{H_u}^2}.
\label{eq:tan-2beta}
\end{equation}
We are interested in the region where $\tan\beta$ is not large in order 
for the first term in Eq.~(\ref{eq:quartic-model-2}) to be effective. 
For $\lambda \simeq 0.8$, for example, we obtain a sufficiently 
large Higgs boson mass for $\tan\beta \simlt (2\!\sim\!3)$, depending 
on the values of $\epsilon, \epsilon'$ and the size of the one-loop 
top-stop correction to the Higgs quartic coupling. 

An approximate formula for the fine-tuning parameter can be 
obtained by inspecting the sum of the two minimization equations, 
Eqs.~(\ref{eq:minimize-1-Ms},~\ref{eq:minimize-2-Ms}):
\begin{equation}
  v^2 \left( \frac{2 \lambda^2 \mu}{M_S} \sin 2\beta 
    + \frac{2 \lambda^2 \mu}{M_S \sin 2\beta} 
    - \frac{4 \lambda^2 \mu^2}{M_S^2} \right)
  = \frac{2 \mu B}{\sin 2\beta} - (2 \mu^2 + m_{H_u}^2 + m_{H_d}^2).
\label{eq:determine-v}
\end{equation}
The combination $2\mu^2+m_{H_u}^2+m_{H_d}^2$ is positive in our 
scenario, so that the electroweak scale is determined by the balance 
between this combination and the first term in the right-hand-side. 
We also note that in the parameter region we are interested, e.g. 
$\tan\beta \simlt 2$, the combination inside the parenthesis in the 
left-hand-side is approximately $\lambda_h$.  The fine-tuning parameter is 
then given by the ratio between $\lambda_h v^2$ and $2 \mu B/\sin 2\beta$, 
where the latter quantity is approximately the charged Higgs boson mass 
squared for $\mu B$ sufficiently larger than $\lambda^2 v^2 \mu/M_S$. 
We thus obtain
\begin{equation}
  \Delta^{-1} \approx \frac{M_{\rm Higgs}^2}{M_{H^{\pm}}^2},
\label{eq:fine-tuning-2}
\end{equation}
which provides a rough measure for the fine-tuning in the present 
scenario.  Since the formula is approximately the same as that of 
the previous scenario, Eq.~(\ref{eq:fine-tuning-1}), the implications 
on $M_{H^{\pm}}^2$ in terms of $M_{\rm Higgs}$ are almost identical.

We note that this scenario is exactly the one employed in 
Ref.~\cite{NPT1} to reduce the fine-tuning, where the supersymmetric 
masses for $S$ and the Higgs doublets, as well as the holomorphic 
supersymmetry breaking masses for these fields, are obtained through 
supergravity effects.  In fact, one can check that our approximate 
formula in Eq.~(\ref{eq:fine-tuning-2}) well estimates the values 
of the fine-tuning parameter $\tilde{\Delta}^{-1}$ calculated in 
that paper.

\subsection{Implications}
\label{subsec:impli}

One of the solid implications of our scenario is that $\tan\beta$ is 
expected to be small, e.g.
\begin{equation}
  \tan\beta \simlt 3.
\label{eq:tan-beta}
\end{equation}
In both scenarios discussed in the previous two subsections, obtaining 
a large $\tan\beta$ requires fine-tuning.  The small $\tan\beta$ is 
also required to push up the Higgs boson mass using the quartic coupling 
arising from the $F$-term potential.

An important feature of the present scenario is that small top squark 
masses are allowed because they do not play a major role in raising the 
Higgs boson mass nor in triggering electroweak symmetry breaking.  In 
fact, in order for our scenario to be relevant for the solution to the 
supersymmetric fine-tuning problem, the top squarks should be light. 
How light should they be?  Let us look at Eq.~(\ref{eq:min-correct-2}) 
for the case with a large $m_S^2$.  To avoid fine-tuning, each contribution 
to the right-hand-side must not be much larger than the left-hand-side. 
Now, for small $\tan\beta$, the quantity in the left-hand-side is roughly 
the square of the physical Higgs boson mass $M_{\rm Higgs}^2$ (see 
Eqs.~(\ref{eq:higgs-mass},~\ref{eq:quartic-model-1})).  We thus have 
a rough upper bound on the size of the top-Yukawa contribution to the 
Higgs mass-squared parameter
\begin{equation}
  |\delta m_h^2| \simlt \frac{M_{\rm Higgs}^2}{\Delta^{-1}},
\label{eq:bound-top-contr}
\end{equation}
where $\Delta^{-1}$ is the amount of fine-tuning.  This in turn gives an 
upper bound on the top squark masses through Eq.~(\ref{eq:corr-Higgs}). 
Requiring $\Delta^{-1} \geq 20\%$, for example, we obtain
\begin{equation}
  m_{\tilde{t}} \simlt 720~{\rm GeV} 
    \biggl( \frac{M_{\rm Higgs}}{150~{\rm GeV}} \biggr)
    \biggl( \frac{3.5}{\ln(M_{\rm mess}/m_{\tilde{t}})} \biggr)^{1/2},
\label{eq:bound-stop}
\end{equation}
where $M_{\rm mess}$ is the scale at which the squark and slepton masses 
are generated.  This is not an extremely strong bound.  (The bound can 
become even weaker if $\tan\beta$ is appreciably larger than $1$; see 
Eq.~(\ref{eq:quartic-model-1})).  The bound in Eq.~(\ref{eq:bound-stop}) 
implies that our scenario can evade fine-tuning for a broad range of 
the top-squark masses.  In fact, in most cases the limiting factor 
for fine-tuning is the mass of the charged Higgs boson, rather than 
the top squark masses (see e.g. Eq.~(\ref{eq:fine-tuning-1})).  The 
situation is quite similar in the case with a large $M_S$ discussed in 
section~\ref{subsec:large-MS}.  We obtain Eqs.~(\ref{eq:bound-top-contr},%
~\ref{eq:bound-stop}) from Eqs.~(\ref{eq:determine-v},~%
\ref{eq:quartic-model-2}) in an exactly analogous manner.

It is interesting to compare the expression in Eq.~(\ref{eq:bound-top-contr}) 
with the corresponding one in the MSSM.  In the MSSM, the electroweak VEV 
is determined by Eq.~(\ref{eq:min-wrong-1}) with $\epsilon$ and $\epsilon'$ 
set to $1$.  For a reasonably large $\tan\beta$, this equation can be 
approximately written as $M_{\rm Higgs}^2/2 = - m_{H_u}^2 - |\mu|^2$. 
We then obtain the bound on the top-Yukawa contribution
\begin{equation}
  |\delta m_h^2| \simlt \frac{M_{\rm Higgs}^2}{2 \Delta^{-1}},
\label{eq:bound-top-contr-MSSM}
\end{equation}
which is a factor $2$ tighter than that in Eq.~(\ref{eq:bound-top-contr}). 
This shows that, even disregarding the issue of the LEP~II bound on 
the Higgs boson mass, the quartic coupling from the $F$-term has some 
advantage over those from the $SU(2)_L \times U(1)_Y$ $D$-terms.  For 
a fixed value of $\Delta^{-1}$, theories with the $F$-term quartic 
coupling can accommodate larger values for the top squark masses than 
theories with the $D$-term quartic couplings.

\section{Conclusions}
\label{sec:concl}

In this paper we have considered a scenario in which the dominant 
stabilizing force in the Higgs potential arises from the $F$-term 
potential, rather than the conventional $SU(2)_L \times U(1)_Y$ $D$-terms, 
in supersymmetric theories.  The required quartic coupling arises from the 
superpotential interaction between the Higgs doublets and a singlet field, 
but unlike in the case of the next-to-minimal supersymmetric standard 
model the singlet field is not responsible for generating the $\mu$ 
and $\mu B$ terms for the Higgs doublets.  This allows a larger freedom 
for the superpotential and supersymmetry-breaking terms for the singlet 
field than in the case where $\mu$ and $\mu B$ arise from the VEV of 
the singlet field.

To use the $F$-term quartic coupling as the dominant stabilizing force, 
the mass-squared parameters for the up- and down-type Higgs doublets 
should both be positive.  This implies that electroweak symmetry breaking 
is triggered by the holomorphic supersymmetry breaking mass term for 
the Higgs doublets, the $\mu B$ term.  The scenario, therefore, requires 
deviations from the conventional picture not only in the stabilizing 
force for the Higgs potential but also in the origin of electroweak 
symmetry breaking.  In this scenario, the role of the top-stop loop 
correction to the up-type Higgs doublet is to make $\tan\beta$ deviate 
from $1$, besides facilitating electroweak symmetry breaking.

One of the advantages of the present scenario is that the masses of 
the top squarks can be small as they do not play a major role in raising 
the Higgs boson mass or in triggering electroweak symmetry breaking. 
This makes it possible to reduce or eliminate the fine-tuning problem 
in supersymmetric theories.  We have studied electroweak symmetry 
breaking in the limit where the dynamics of the singlet field essentially 
decouples.  There are two such cases, in which the decoupling occurs 
either due to the supersymmetry-breaking or supersymmetric mass for the 
singlet field.  We have found that fine-tuning can be significantly 
reduced in both cases.  The fine-tuning parameter $\Delta^{-1}$ is given 
by the squared ratio of the lightest Higgs boson mass to the charged 
Higgs boson mass, which can be made better than $20\%$ for the charged 
Higgs boson mass smaller than about $300~{\rm GeV}$.

Our scenario can be combined with any supersymmetry breaking models 
in which relatively small top squark masses are generated at 
a low scale, such as the one in~\cite{Nomura:2004is}, to reduce 
fine-tuning. Examples for such theories are given, for example, 
in Refs.~\cite{Chacko:2005ra,NPT1}.  Generic predictions for these 
theories include a small $\tan\beta$, $\tan\beta \simlt 3$, relatively 
light top squarks, $m_{\tilde{t}} \simlt 720~{\rm GeV} (M_{\rm Higgs}%
/150~{\rm GeV}) (3.5/\ln(M_{\rm mess}/m_{\tilde{t}}))^{1/2}$, and 
the existence of the states for a singlet chiral supermultiplet below 
about a TeV.  It will be interesting to further explore possible 
models in which the present mechanism is incorporated and the 
fine-tuning of electroweak symmetry breaking is reduced to the level 
discussed in this paper.

\section*{Acknowledgments}

Y.N. thanks David Tucker-Smith for conversation. 
This work was supported in part by the Director, Office of Science, Office 
of High Energy and Nuclear Physics, of the US Department of Energy under 
Contract DE-AC02-05CH11231.  The work of Y.N. was also supported by the 
National Science Foundation under grant PHY-0403380, by a DOE Outstanding 
Junior Investigator award, and by an Alfred P. Sloan Research Fellowship.

\section*{Appendix}

In this appendix, we integrate out the $S$ supermultiplet in the general 
setup and obtain the effective potential for the two Higgs doublets. 
Our starting point is the interactions given by Eqs.~(\ref{eq:superpot},%
~\ref{eq:soft-Lag}).  We assume that the linear $S$ terms in $\hat{W}(S)$ 
and $\hat{\cal L}_{\rm soft}(S,H_u,H_d)$ are absent, since their effects 
can be absorbed in the appropriate redefinitions of the other parameters 
and the origin of the $S$ field. 

The relevant terms in $\hat{W}(S)$ and $\hat{\cal L}_{\rm soft}(S,H_u,H_d)$ 
are given by
\begin{eqnarray}
  \hat{W}(S) &=& \frac{M_S}{2} S^2,
\\
  \hat{\cal L}_{\rm soft}(S,H_u,H_d) &=& -m_S^2 |S|^2 
    - \left( \frac{b_S}{2} S^2 + a_\lambda S H_u H_d + {\rm h.c.} \right),
\end{eqnarray}
where we have neglected the $S^3$ term in $\hat{W}(S)$, which has 
negligible effects for small $\langle S \rangle$.  The interactions between 
the bosonic fields in $S$, $H_u$ and $H_d$ are given in components by
\begin{equation}
  {\cal L}_{\rm int.} = F_S (\lambda H_u H_d) 
    + S (\lambda F_{H_u} H_d + \lambda H_u F_{H_d} - a_\lambda H_u H_d) 
    + {\rm h.c.},
\end{equation}
where we have used the same symbol for a chiral superfield and its scalar 
component.  The effective Lagrangian is obtained by integrating out the 
$S$ and $F_S$ fields using their propagators:
\begin{equation}
  G = \left( \begin{array}{cc|cc} 
    G_{F_S F_S^\dagger} & G_{F_S F_S} & 
    G_{F_S S^\dagger} & G_{F_S S} \\
    G_{F_S^\dagger F_S^\dagger} & G_{F_S^\dagger F_S} & 
    G_{F_S^\dagger S^\dagger} & G_{F_S^\dagger S} \\ \hline
    G_{S F_S^\dagger} & G_{S F_S} & 
    G_{S S^\dagger} & G_{S S} \\
    G_{S^\dagger F_S^\dagger} & G_{S^\dagger F_S} & 
    G_{S^\dagger S^\dagger} & G_{S^\dagger S} 
  \end{array} \right)
  \equiv \frac{i}{-(p^2+|M_S|^2 + m_S^2)^2 + |b_S|^2} 
  \left( \begin{array}{c|c} 
    \hat{G}_{FF} & \hat{G}_{FS} \\ \hline
    \hat{G}_{FS}^\dagger & \hat{G}_{SS} 
  \end{array} \right),
\label{eq:S-prop}
\end{equation}
where
\begin{eqnarray}
  && \hat{G}_{FF} = \left( \begin{array}{cc} 
    -(p^2+m_S^2)(p^2+|M_S|^2+m_S^2)+|b_S|^2 & -b_S M_S^{*2} \\ 
    -b_S^* M_S^2 & -(p^2+m_S^2)(p^2+|M_S|^2+m_S^2)+|b_S|^2
  \end{array} \right),
\nonumber\\
  && \hat{G}_{FS} = \left( \begin{array}{cc} 
    b_S M_S^* & -M_S^* (p^2+|M_S|^2+m_S^2) \\ 
    -M_S (p^2+|M_S|^2+m_S^2) & b_S^* M_S 
  \end{array} \right),
\nonumber\\
  && \hat{G}_{SS} = \left( \begin{array}{cc} 
    p^2+|M_S|^2+m_S^2 & -b_S^* \\ 
    -b_S & p^2+|M_S|^2+m_S^2
  \end{array} \right).
\nonumber
\end{eqnarray}
Here, $p^2 = p^\mu p_\mu$ is the invariant 4-momentum, and the metric 
is given by $\eta_{\mu\nu} = {\rm diag}(-,+,+,+)$.

Neglecting the corrections to the wavefunction renormalizations, the 
effective potential for the Higgs doublets are obtained using the 
$S$ and $F_S$ propagators at zero momentum:
\begin{eqnarray}
  V &=& \frac{m_S^2(|M_S|^2+m_S^2)-|b_S|^2}{(|M_S|^2+m_S^2)^2-|b_S|^2}
    |\lambda H_u H_d|^2
\nonumber\\
  && {} - \frac{|M_S|^2+m_S^2}{(|M_S|^2+m_S^2)^2-|b_S|^2}
    |\lambda F_{H_u} H_d + \lambda H_u F_{H_d} - a_\lambda H_u H_d|^2
\nonumber\\
  && {} + \frac{1}{(|M_S|^2+m_S^2)^2-|b_S|^2} \Biggl[ 
    \frac{b_S M_S^{*2}}{2} (\lambda H_u H_d)^2
    + \frac{b_S^*}{2}(\lambda F_{H_u} H_d 
      + \lambda H_u F_{H_d} - a_\lambda H_u H_d)^2
\nonumber\\
  && {} + M_S^*(|M_S|^2+m_S^2) (\lambda H_u H_d) 
      (\lambda F_{H_u} H_d + \lambda H_u F_{H_d} - a_\lambda H_u H_d) 
\nonumber\\
  && {} - b_S^* M_S (\lambda H_u H_d)^\dagger 
      (\lambda F_{H_u} H_d + \lambda H_u F_{H_d} - a_\lambda H_u H_d) 
    + {\rm h.c.} \Biggr].
\label{eq:app-pot}
\end{eqnarray}
While we do not present the explicit expression, it is also easy 
to obtain the effective Lagrangian for the Higgs doublets using the 
full momentum-dependent propagators of Eq.~(\ref{eq:S-prop}).

\end{document}